\documentclass[12pt]{article}
\usepackage{graphicx}

\def\hybrid{\topmargin 0pt      \oddsidemargin 0pt
        \headheight 0pt \headsep 0pt
       \voffset-1cm
        \textwidth 6.25in       
       \textheight 9.5in       
        \marginparwidth 0.0in
        \parskip 5pt plus 1pt   \jot = 1.5ex}
\catcode`\@=11
\def\marginnote#1{}

\newcount\hour
\newcount\minute
\newtoks\amorpm
\hour=\time\divide\hour by60
\minute=\time{\multiply\hour by60 \global\advance\minute by-\hour}
\edef\standardtime{{\ifnum\hour<12 \global\amorpm={am}%
        \else\global\amorpm={pm}\advance\hour by-12 \fi
        \ifnum\hour=0 \hour=12 \fi
        \number\hour:\ifnum\minute<10 0\fi\number\minute\the\amorpm}}
\edef\militarytime{\number\hour:\ifnum\minute<10 0\fi\number\minute}

\def\draftlabel#1{{\@bsphack\if@filesw {\let\thepage\relax
   \xdef\@gtempa{\write\@auxout{\string
      \newlabel{#1}{{\@currentlabel}{\thepage}}}}}\@gtempa
   \if@nobreak \ifvmode\nobreak\fi\fi\fi\@esphack}
        \gdef\@eqnlabel{#1}}
\def\@eqnlabel{}
\def\@vacuum{}
\def\draftmarginnote#1{\marginpar{\raggedright\scriptsize\tt#1}}

\def\draftlabel#1{{\@bsphack\if@filesw {\let\thepage\relax
   \xdef\@gtempa{\write\@auxout{\string
      \newlabel{#1}{{\@currentlabel}{\thepage}}}}}\@gtempa
   \if@nobreak \ifvmode\nobreak\fi\fi\fi\@esphack}
        \gdef\@eqnlabel{#1}}
\def\@eqnlabel{}
\def\@vacuum{}
\def\draftmarginnote#1{\marginpar{\raggedright\scriptsize\tt#1}}

\def\draft{\oddsidemargin -.5truein
        \def\@oddfoot{\sl preliminary draft \hfil
        \rm\thepage\hfil\sl\today\quad\militarytime}
        \let\@evenfoot\@oddfoot \overfullrule 3pt
        \let\label=\draftlabel
        \let\marginnote=\draftmarginnote
   \def\@eqnnum{(\theequation)\rlap{\kern\marginparsep\tt\@eqnlabel}%
\global\let\@eqnlabel\@vacuum}  }

\def\numberbysection{\@addtoreset{equation}{section}
        \def\theequation{\thesection.\arabic{equation}}}

\def\underline#1{\relax\ifmmode\@@underline#1\else
        $\@@underline{\hbox{#1}}$\relax\fi}

\def\titlepage{\@restonecolfalse\if@twocolumn\@restonecoltrue\onecolumn
     \else \newpage \fi \thispagestyle{empty}\c@page\z@
        \def\thefootnote{\fnsymbol{footnote}} }

\def\endtitlepage{\if@restonecol\twocolumn \else  \fi
        \def\thefootnote{\arabic{footnote}}
        \setcounter{footnote}{0}}  
\relax

\hybrid

\newfont{\Bbb}{msbm10 scaled 1\@ptsize00}
\newfont{\Bbbb}{msbm7 scaled 1\@ptsize00}
\newcommand{\CC}{\mbox{\Bbb C}}

\newcommand{\ccc}{\raise-1pt\hbox{$\mbox{\Bbbb C}$}}

\newcommand{\DDD}{\raise-1pt\hbox{$\mbox{\Bbbb D}$}}

\newcommand{\UUU}{\raise-1pt\hbox{$\mbox{\Bbbb U}$}}

\newcommand{\ZZ}{\mbox{\Bbb Z}}
\newcommand{\z}{\raise-1pt\hbox{$\mbox{\Bbbb Z}$}}

\newcommand{\SSS}{\mbox{\Bbb S}}
\newcommand{\sss}{\raise-1pt\hbox{$\mbox{\Bbbb S}$}}

\def\beq{\begin{equation}}
\def\eeq{\end{equation}}
\def\p{\partial}

\newtheorem{lemma-definition}{Lemma-Definition}[section]

\begin{document}

\begin{titlepage}

\title{Elliptic families of solutions to constrained Toda hierarchy}

\author{A.~Zabrodin\thanks{
Skolkovo Institute of Science and Technology, 143026, Moscow, Russia and
National Research University Higher School of Economics,
20 Myasnitskaya Ulitsa,
Moscow 101000, Russia and
ITEP NRC KI, 25
B.Cheremushkinskaya, Moscow 117218, Russia;
e-mail: zabrodin@itep.ru}}

\date{January 2022}
\maketitle

\vspace{-7cm} \centerline{ \hfill ITEP-TH-01/22}\vspace{7cm}

\vspace{-1.5cm}

\begin{center}

{\it To Andrey K. Pogrebkov on his 75th birthday}

\end{center}

\begin{abstract}

We study elliptic families of solutions to the recently introduced 
constrained Toda hierarchy, i.e., solutions which are
elliptic functions of some linear combination of the hierarchical times. 
Equations of motion for poles of such solutions are obtained.

\end{abstract}

\end{titlepage}

\vspace{5mm}

%

\tableofcontents

\vspace{5mm}

\section{Introduction}

The Toda lattice hierarchy \cite{UT84} is an 
infinite set of evolution equations
for two Lax operators $L$, $\bar L$ which are pseudo-difference operators
in the variable $x$.
Let $\{t_k\}_{k\in \z}$ be the infinite set of independent variables
(times) indexed by integer numbers.  
The hierarchy is defined by the infinite set of Lax equations 
(evolution equations 
for the Lax operators in the times $t_k$). 
They are equivalent to differential-difference equations for 
the coefficient functions of the Lax operators. These equations 
are differential with respect to the times $t_k$ with $k\neq 0$ and
difference with respect to the time $t_0=x/\eta$ ($\eta$ plays the role
of the lattice spacing). 
An equivalent formulation is via the tau-function 
$\tau$ \cite{DJKM83,JM83} which is a function of the 
infinite set of independent variables satisfying bilinear 
functional relations.

The constrained Toda hierarchy was recently introduced in \cite{KZ21}. 
It is a subhierarchy of the Toda lattice hierarchy defined by the 
constraint 
\beq\label{int1}
\bar L =L^{\dag}
\eeq
(in the symmetric gauge).
As is shown in \cite{KZ21}, 
the constraint is preserved by the flows $\p_{t_k}-\p_{t_{-k}}$ and is destroyed
by the flows $\p_{t_k}+\p_{t_{-k}}$, so one has to
put $t_k +t_{-k}=0$. 

The investigation of dynamics of poles of singular 
solutions to nonlinear integrable
equations was initiated in the seminal paper \cite{AMM77}, 
where elliptic and rational
solutions to the Korteweg-de Vries and Boussinesq equations were studied.
As it was proved later in \cite{Krichever78,CC77}, poles of solutions 
to the Kadomtsev-Petviashvili (KP) equation 
which are rational functions of $t_1$, as functions of
the second hierarchical time $t_2$, move as particles of the integrable
Calogero-Moser system \cite{Calogero71,Calogero75,Moser75,OP81}.
This correspondence was extended to elliptic solutions in \cite{Krichever80}.
Dynamics of poles of elliptic solutions to the 2D Toda lattice and
modified KP (mKP) equations was
studied in \cite{KZ95}, see also \cite{Z19}.
It was proved that the poles move as particles of the
integrable Ruijsenaars-Schneider many-body system \cite{RS86,Ruijs12}
which is a relativistic generalization
of the Calogero-Moser system.

The study of more general {\it elliptic families} of solutions to 
nonlinear integrable hierarchies, i.e. solutions which are elliptic 
functions of a general linear combination of the higher times
of the hierarchy, was initiated
in \cite{AKV02}. It was shown that poles of such 
solutions as functions of $t_1$ and $t_2$ move
according to equations of motion of the field generalization of the 
Calogero-Moser system. Recently, similar results for elliptic families 
of solutions to the Toda hierarchy were obtained in \cite{ZZ21}, where
the field generalization of the Ruijsenaars-Schneider model was 
introduced.

The aim of this paper is to study elliptic families of 
solutions to the constrained
Toda hierarchy. We will derive equations of motion for poles of
these solutions. This system can be regarded as a field generalization of
the system obtained in \cite{KZ21}.

\section{Constrained Toda hierarchy}

\subsection{The Toda hierarhy}

We begin with the Toda lattice hierarchy \cite{UT84} in the symmetric gauge
\cite{Takebe1,Takebe2}. The two Lax operators are 
the pseudo-difference operators
\beq\label{t1}
L=c(x)e^{\eta \p _x}+\sum_{k\geq 0}U_k(x)e^{-k\eta \p_x}, \quad
\bar L=c(x-\eta )e^{-\eta \p _x}+\sum_{k\geq 0}\bar U_k(x)e^{k\eta \p_x},
\eeq
where $e^{k\eta \p_x}$ are shift operators acting on functions 
of $x$ as $e^{k\eta \p_x}f(n)=f(x+k\eta )$.
Given the Lax operators, one can introduce the difference operators
\beq\label{t2}
B_m =(L^m)_{>0} +\frac{1}{2} (L^m)_{0}, \quad
B_{-m}=(\bar L^m)_{<0}+\frac{1}{2} (\bar L^m)_{0}, \quad m=1,2,3, \ldots ,
\eeq
where for a subset $\SSS \subset \ZZ$, we denote
$\displaystyle{\Bigl (\sum_{k\in \z} U_k e^{k \p_n}\Bigr )_{\sss}=
\sum_{k\in \sss} U_k e^{k \p_n}}$. The Toda lattice hierarchy is given by the 
Lax equations
\beq\label{t3}
\p_{t_m}L=[B_m, L], \quad \p_{t_m}\bar L=[B_m, \bar L]
\eeq
which define the hierarchical flows parametrized by the 
times $t_m$ for any non-zero
integer $m$. An equivalent formulation is through 
the zero curvature (Zakharov-Shabat)
equations
\beq\label{t4}
\p_{t_k}B_m-\p_{t_m}B_k +[B_m, B_k]=0.
\eeq

One of the main objects related to the hierarchy is 
the tau-function $\tau (x, {\bf t})$ which we denote below simply as
$\tau (x)$ skipping the dependence on the times. 
The coefficient $c(x)$ is expressed through 
the tau-function by the formula
\beq\label{t11}
c(x)=\left (\frac{\tau (x+2\eta )\tau (x)}{\tau^2 (x+\eta )}\right )^{1/2}.
\eeq

\subsection{Specialization to the constrained Toda hierarchy}

The constrained Toda hierarchy is obtained by imposing the constraint
\beq\label{t5}
\bar L =L^{\dag},
\eeq
where the ${}^\dag$-operation is defined as $(f(x)\circ e^{k\eta 
\p_x})^{\dag}
=e^{-k\eta \p_x}\circ f(x)$. This implies that $\bar U_k(x)=U_k(x+k\eta )$. 
It is easy to see that
the constraint is preserved by the flows
$\p_{t_k}-\p_{t_{-k}}$ and is destroyed
by the flows $\p_{t_k}+\p_{t_{-k}}$, so one has to
put $t_k +t_{-k}=0$. In this way all the coefficient functions can be regarded as functions of
$t_k$ with $k>0$ (and of $x$) only. Introducing difference operators
\beq\label{t6}
A_m =B_m-B_{-m},
\eeq
we can write the Lax and Zakharov-Shabat equations of the constrained hierarchy
in the form
\beq\label{t7}
\p_{t_m}L=[A_m, L], \quad [\p_{t_m}-A_m, \, \p_{t_k}-A_k]=0, \quad m>0.
\eeq
In particular,
$$
A_1=c(x)e^{\eta \p_x}-c(x-\eta )e^{-\eta \p_x},
$$
$$
A_2=c(x)c(x+\eta )e^{2\eta \p_x}+c(x)(v(x)+v(x+\eta ))e^{\eta \p_x}
$$
$$
-c(x-\eta )(v(x)+v(x-\eta ))e^{-\eta \p_x}-
c(x-\eta )c(x-2\eta )e^{-2\eta \p_x},
$$
where $v(x)=U_0(x)$.

The Zakharov-Shabat and Lax equations are compatibility conditions for the 
linear problems
\beq\label{t8}
\p_{t_m}\psi =A_m\psi , \quad L\psi =z\psi 
\eeq
for the wave function $\psi =\psi (x,{\bf t}, z )$ 
depending on the spectral parameter $z\in \CC$
(and on all the times ${\bf t}=\{t_1, t_2, t_3 , \ldots \}$).
In particular, we have the linear problem
\beq\label{t10}
\p_{t_1}\psi (x)=c(x)\psi (x+\eta )-c(x-\eta )\psi (x-\eta ),
\eeq
where we do not indicate the dependence on ${\bf t}$ for brevity.

Introducing the wave function $\Psi (x)$ by means of the relation
\beq\label{t12}
\Psi (x)=\left (\frac{\tau (x+\eta )}{\tau (x)}\right )^{1/2}\psi (x),
\eeq
we represent the linear problem (\ref{t10}) in the form
\beq\label{t13}
\p_{t_1}\Psi (x)=\Psi (x+\eta )+b(x)\Psi (x) -a(x)\Psi (x-\eta ),
\eeq
where
\beq\label{t14}
b(x)=\frac{1}{2}\, \p_{t_1}\log \frac{\tau (x+\eta )}{\tau (x)},
\quad
a(x)=\frac{\tau (x+\eta )\tau (x-\eta )}{\tau^2 (x)}.
\eeq

\section{Elliptic families}

\subsection{Elliptic families among general 
algebraic-geometrical solutions}

We are going to consider solutions
that are elliptic functions of
a linear combination
$\displaystyle{\lambda =\beta_0x+\sum_k \beta_k t_k}$ of 
higher times of the hierarchy.
We call them elliptic families. 
The elliptic families
form a particular class of algebraic-geometrical 
solutions associated with an algebraic curve
$\Gamma$ of genus $g$ with some additional data. 
An algebraic-geometrical solution is said to be elliptic
with respect to some variable $\lambda$ if there exists a $g$-dimensional
vector ${\bf W}$ such that it spans an elliptic curve ${\cal E}$
embedded in the Jacobian
of the curve $\Gamma$. The tau-function of such solution has the form
\beq\label{toda8aa}
\tau (x, {\bf t}, \lambda )=e^{Q(x, {\bf t})}
\Theta \Bigl (
{\bf V}_0 x/\eta +\sum_{k\geq 1}{\bf V}_k t_k +{\bf W}\lambda
+{\bf Z}\Bigr ),
\eeq
where $\Theta$ is the Riemann theta-function with the Riemann 
matrix being the matrix of $b$-periods of 
normalized holomorphic differentials on $\Gamma$,
and $Q(x, {\bf t})$ 
is a quadratic form in the variable $x$ and the 
hierarchical times ${\bf t}=\{t_1, t_2, t_3, \ldots \}$. 
The vectors ${\bf V}_k$ are related to
$b$-periods of certain normalized meromorphic differentials on $\Gamma$.
The existence of a $g$-dimensional
vector ${\bf W}$ such that it spans an elliptic curve ${\cal E}$
embedded in the Jacobian
is a nontrivial transcendental condition.
If such a vector ${\bf W}$ exists, then the theta-divisor intersects
the shifted elliptic curve $\displaystyle{{\cal E}+{\bf V}_0 x/\eta +
\sum_k {\bf V}_k t_k}$ at
a finite number of points $\lambda_i =\lambda_i(x,{\bf t})$. Therefore,
for elliptic families we can write:
\beq\label{toda8bb}
\Theta \Bigl (
{\bf V}_0 x/\eta +\sum_{k\geq 1}{\bf V}_k t_k +{\bf W}\lambda
+{\bf Z}\Bigr )=f(x, {\bf t})e^{\gamma_1\lambda +\gamma_2\lambda^2}\prod _{i=1}^N
\sigma (\lambda -\lambda_i(x, {\bf t}))
\eeq
with a function $f(x, {\bf t})$ and
some constants $\gamma_1, \gamma_2$. Here $\sigma (\lambda )$ is the
Weierstrass $\sigma$-function 
with quasi-periods $2\omega_1$, $2\omega_2$ (such that
${\rm Im} (\omega_2/ \omega_1 )>0$)
defined
by the infinite product
\beq\label{w01}
\sigma (x)=\sigma (x |\, \omega_1 , \omega_2)=
x\prod_{s\neq 0}\Bigl (1-\frac{x}{s}\Bigr )\, e^{\frac{x}{s}+\frac{x^2}{2s^2}},
\quad s=2\omega_1 m_1+2\omega_2 m_2 \quad \mbox{with integer $m_1, m_2$}.
\eeq
Below we also use the Weierstrass $\zeta$-function 
$\zeta (x)=\sigma '(x)/\sigma (x)$. 
The form of the exponential
factor in the right hand side
of (\ref{toda8bb}) follows from monodromy properties of the theta-function.
The zeros $\lambda_i$ of the
tau-function are poles of the elliptic solutions.

From (\ref{t14}), (\ref{toda8aa}), (\ref{toda8bb}) we conclude that
for an elliptic family the coefficients
$b(x)=b(x, \lambda )$, $a(x)=a(x, \lambda )$ have the form
\beq\label{el1}
b(x,\lambda )=\frac{1}{2}\sum_{i=1}^N 
\Bigl (\dot \lambda_i (x)\zeta (\lambda -\lambda_i(x))
-\dot \lambda_i (x+\eta )\zeta (\lambda -\lambda_i(x+\eta ))\Bigr ) +c(x,t),
\eeq
\beq\label{el2}
a(x, \lambda )=g(x, t)\prod_{i=1}^N \frac{\sigma (\lambda -
\lambda_i(x+\eta ))\sigma (\lambda -\lambda_i(x-\eta ))}{\sigma^2
(\lambda -\lambda_i(x))},
\eeq
where dot means the $t_1$-derivative and $c(x,t)$, $g(x, t)$ 
are some functions.

\subsection{Double-Bloch functions}

Our strategy is to find
$b(x,\lambda )$, $a(x,\lambda )$ such that the equation (\ref{t13})
has sufficiently many double-Bloch solutions. 
The existence of double-Bloch solutions turn out to be
a rather restrictive condition.

A meromorphic function $f(\lambda )$ 
is called a {\it double-Bloch} function if it satisfies the
following monodromy properties:
\beq\label{bl1}
f(\lambda +2\omega_{\alpha})=B_{\alpha}f(\lambda ), \quad \alpha =1,2.
\eeq
The complex constants $B_{\alpha}$ are called Bloch multipliers. 
Let the 
function $\Phi (\lambda , z)$ be defined by
\beq\label{w02}
\Phi (\lambda , z )=\frac{\sigma (\lambda +z )}{\sigma (z )\sigma (\lambda )}\,
e^{-\zeta (z )\lambda }.
\eeq
It has a simple pole
at $\lambda =0$ with residue $1$. 
The quasiperiodicity properties of the function $\Phi$ in the variable $\lambda $ 
are
\beq\label{bloch}
\Phi (\lambda +2\omega_{\alpha} , z )=e^{2(\zeta (\omega_{\alpha} )z -
\zeta (z )\omega_{\alpha} )}
\Phi (\lambda , z ),
\eeq
so it can be regarded as an elementary double-Bloch function having only one 
pole. The variable $z$ has the meaning of the spectral parameter.
As a function of $z$, $\Phi (\lambda , z)$ is a double-periodic function.
In what follows we often suppress the second argument of 
$\Phi$ writing simply
$\Phi (\lambda , z):=\Phi (\lambda )$.

The double-Bloch functions with several simple 
poles $\lambda_i$ can be represented in the form
\beq\label{el3}
\Psi (\lambda )=\sum_i c_i(x)\Phi (\lambda -\lambda_i(x), z ),
\eeq
where $c_i$ are residues at the poles $\lambda_i$

\subsection{Derivation of the equations of motion}

In order to 
derive equations of motion for the zeros of
the tau-function $\lambda_i(x)$, 
we substitute (\ref{el3}), (\ref{el1}), (\ref{el2}) into the linear
problem (\ref{t13}) and cancel the poles. 
The substitution gives:
$$
\sum_i \dot c_i(x)\Phi (\lambda -\lambda_i(x))-\sum_i c_i(x)\dot \lambda_i(x)
\Phi ' (\lambda -\lambda_i(x))-\sum_i c_i(x+\eta )
\Phi (\lambda -\lambda_i(x+\eta ))
$$
$$
-\frac{1}{2}
\sum_i \Bigl ((\dot \lambda_i (x)\zeta (\lambda -\lambda_i(x))
-\dot \lambda_i (x+\eta )\zeta (\lambda -\lambda_i(x+\eta ))\Bigr ) \sum_j
c_j(x)\Phi (\lambda -\lambda_j(x))
$$
$$
-c(x,t)\sum_i c_i(x)\Phi (\lambda -\lambda_i(x))
$$
$$
+g(x,t)\prod_i 
\frac{\sigma (\lambda -
\lambda_i(x+\eta ))\sigma (\lambda -\lambda_i(x-\eta ))}{\sigma^2
(\lambda -\lambda_i(x))}\sum_j c_j(x-\eta )
\Phi (\lambda -\lambda_j(x-\eta ))=0.
$$
The cancellation of simple poles at $\lambda =\lambda_i(x+\eta )$
yields the equation
\beq\label{el4}
c_i(x+\eta )=\frac{1}{2}\dot \lambda_i(x+\eta )\sum_j c_j(x)
\Phi (\lambda_i(x+\eta)-\lambda_j(x)).
\eeq
The cancellation of double poles at $\lambda =\lambda_i(x)$
yields the equation
\beq\label{el5}
\begin{array}{l}
\displaystyle{
c_i(x+\eta )=-2\, \frac{g(x+\eta , t)}{\dot \lambda_i(x+\eta )}\,
\frac{\prod\limits_{l}\sigma (\lambda_i(x+\eta )-\lambda_l(x+2\eta ))
\sigma (\lambda_i(x+\eta )-\lambda_l(x))}{\prod\limits_{l\neq i}
\sigma^2(\lambda_i(x+\eta )-\lambda_l (x+\eta ))}}
\\ \\
\displaystyle{\phantom{aaaaaaaaaaaaaaaaaa}
\times \, \sum_j c_j(x)\Phi (\lambda_i(x+\eta )-\lambda_j(x)).}
\end{array}
\eeq
Comparing (\ref{el4}) and (\ref{el5}), we obtain the following equations
of motion:
\beq\label{el6}
\begin{array}{l}
\dot \lambda_i(x)=2g^{1/2}(x,t)\sigma^{1/2}(\lambda_i(x+\eta )-\lambda_i(x))
\sigma^{1/2}(\lambda_i(x)-\lambda_i(x-\eta ))
\\ \\
\displaystyle{\phantom{aaaaaaaaaaaa}
\times \, \prod_{j\neq i}\frac{\sigma^{1/2}(\lambda_i(x)-\lambda_j(x+\eta ))
\sigma^{1/2}(\lambda_i(x)-\lambda_j(x-\eta ))}{\sigma 
(\lambda_i(x)-\lambda_j(x))}}
\end{array}
\eeq
These are field analogue of the equations of motion 
\beq\label{el6a}
\dot x_i =2\sigma (\eta )\prod_{j\neq i}
\frac{\sigma^{1/2}(x_i-x_j+\eta )
\sigma^{1/2}(x_i-x_j-\eta )}{\sigma 
(x_i-x_j )}
\eeq
from \cite{KZ21}
which are obtained from (\ref{el6}) as a particular case when one sets
$\lambda_i (x)=x+x_i$ (and $g(x,t)=1$). The function $g(x,t)$ in (\ref{el6})
can be fixed by multiplying the equations over $i$ from $1$ to $N$:
\beq\label{el7}
\begin{array}{l}
\displaystyle{
g^{N/2}(x,t)=2^{-N}\prod_{i=1}^N \frac{\dot \lambda_i(x)}{\sigma^{1/2}
(\lambda_i(x+\eta )-\lambda_i(x))
\sigma^{1/2}(\lambda_i(x)-\lambda_i(x-\eta ))}}
\\ \\
\displaystyle{\phantom{aaaaaaaaaaaaa}
\times \, \prod_{i\neq j}\frac{\sigma 
(\lambda_i(x)-\lambda_j(x))}{\sigma^{1/2}(\lambda_i(x)-\lambda_j(x+\eta ))
\sigma^{1/2}(\lambda_i(x)-\lambda_j(x-\eta ))}.
}
\end{array}
\eeq

\subsection{Continuum limit}

The continuum limit is the limit $\eta \to 0$. We write:
$$
\lambda_i(x\pm \eta )=\lambda_i (x)\pm \eta \lambda '(x)+
\frac{\eta^2}{2}\, \lambda ''(x) +O(\eta^3),
$$
where prime means the $x$-derivative. We assume that
$$
g(x,t)=1 +\eta^2 h(x,t)+O(\eta^3).
$$
This assumption is consistent with (\ref{el7}). 
Expanding the right hand side of equations (\ref{el6}), we have:
\beq\label{c1}
\dot \lambda_i=2\eta \lambda_i' \left ( 1+\frac{\eta^2}{2}\, h(x,t)-
\frac{\eta^2}{8} \, \frac{\lambda_i''}{\lambda_i'} -
\frac{\eta^2}{2} \sum_{j\neq i} \Bigl (\lambda_j'' \zeta (\lambda_i-\lambda_j)
+\lambda_j'{}^2 \wp (\lambda_i-\lambda_j)\Bigr )+O(\eta^3)\right ),
\eeq
where $\wp (x)=-\zeta '(x)$ is the Weierstrass $\wp$-function.
The naive $\eta\to 0$ limit is
$$
\dot \lambda_i =2\eta \lambda'_i.
$$
The solution is $\lambda_i =\varphi_i (x+2\eta t)$, where $t=t_1$.
However, a more meaningful limit consists in passing to 
the function $y_i(x,t)$ connected with $\lambda_i(x,t)$ by the 
relation $$\lambda_i(x,t)=y_i(x+2\eta t, \eta^3 t).$$
Then from (\ref{c1}), in the limit $\eta \to 0$, we get the equation
\beq\label{c2}
\dot y_i=\frac{1}{4}\, y_i''-
\sum_{j\neq i} \Bigl (y_i'y_j'' \zeta (y_i-y_j)
+y_i'y_j'{}^2 \wp (y_i-y_j)\Bigr )+y_i'h(x,t),
\eeq
which can be regarded as a field generalization of the equations of motion
for zeros of the CKP tau-function obtained in \cite{KZ21a}. 

\section{Conclusion}

We have studied solutions to the constrained Toda hierarchy which are
elliptic functions of a general linear combination $\lambda$ 
of higher times 
of the hierarchy. For such solutions, the tau-function is essentially 
a product of the Weierstrass sigma-functions of $\lambda$ with 
zeros $\lambda_i$, $i=1, \ldots , N$. We have investigated how
these zeros (poles of the solutions) depend on $x$ and $t_1$
and derived equations of motion (\ref{el6}) for them which are 
first order differential equations in the time $t_1$ and difference 
equations in the space variable $x$. These equations can be regarded 
as a field generalization of the equations of motion for zeros of the
tau-function of the constrained Toda hierarchy obtained in \cite{KZ21}.
The continuum limit of these equations is a field generalization of the
equations of motion for zeros of the CKP tau-function obtained in 
\cite{KZ21a}. It is an open problem to clarify whether these equations
are Hamiltonian. 

\section*{Acknowledgments}

\addcontentsline{toc}{section}{\hspace{6mm}Acknowledgments}

This work has been funded within the framework of the
HSE University Basic Research Program.

\section*{Conflict of Interest}

The author declares that he has no conflicts of interest.

\end{document}